\documentstyle[preprint,aps]{revtex}

\begin{document}
\draft
\preprint{
\begin{tabular}{r}
\text {\bf LPC 95/12}\\
\text {March 1995}\\
\end{tabular}}
\title{
AZIMUTHAL CORRELATIONS IN PHOTON-PHOTON COLLISIONS
}
\author{
                  N.~Arteaga, C.~Carimalo, P.~Kessler, and S.~Ong
        }
\address{Laboratoire de Physique Corpusculaire,
                Coll\`ege de France \\
                11, Place Marcelin Berthelot
                F-75231 Paris Cedex 05, France \\
        }
\author{
                  O.~Panella\footnote{Presently at : Laboratoire de
                  Physique Corpusculaire, Coll\`ege de France, Paris.}
       }
\address{Dipartimento di Fisica, Universit\`a di Perugia
                                          and INFN, Sezione di Perugia\\
                 Via A. Pascoli I-06123 Perugia, Italy  \\
        }
\maketitle

\begin{abstract}
Using the general helicity formula for $\gamma^* \gamma^*$
collisions, we are showing
that it should be possible to determine a number of independent
``structure functions'', i.e. linear combinations
of elements of the two-photon helicity
tensor, through azimuthal correlations in two-body or quasi two-body
reactions induced by the photon-photon interaction, provided certain
experimental conditions are satisfied.
Numerical results of our computations
are presented for some particular processes and dynamic models.

\end{abstract}
\narrowtext 

\section{INTRODUCTION}
\label{sec:intro}

Azimuthal correlations in photon-photon collisions have been studied,
in the past, in a number of papers where, in particular,
the single-tag configuration
was considered \cite{balakin,budnev1,higu,peter,diber,ong1}.
One paper was also devoted to the study of those correlations in a
double-tag configuration with both electrons being tagged at small angle
\cite{ong2}. Let us mention, in addition, a paper\cite{chernyak1}
where the authors investigated azimuthal correlations in pair
production in a no-tag configuration, using the acoplanarity of the
produced particles in the lab frame.

The purpose of the present paper is to show how the potential of
azimuthal correlations, as they can be derived from the general helicity
formula for $\gamma^* \gamma^*$ collisions in the case of 2-body or
quasi 2-body reactions, can
be exploited, in various experimental configurations, in order to extract a
maximum of physical information from measurements of those reactions.
For each of those configurations, we define the corresponding experimental
constraints to be applied.

In section II we write down the general helicity formula as a sum of 13
terms involving each a different dependence on azimuthal angles.
In section III
we apply that formula to four particular experimental configurations where
the analysis of azimuthal correlations should allow one to determine a
number of structure functions $F_i$.
Our treatment of the first two is rather trivial ;
that of the third and fourth one is more sophisticated, since it involves
the use of ``azimuthal selection'',
as we shall define it. Since the
fourth configuration appears to be the most promising one and
has not been
considered elsewhere, we compute, in section IV, a number of corresponding
applications to particular processes and dynamic models.
Section V contains a
brief discussion and conclusion. Kinematic constraints to be applied in
configurations 2 and 4 are being computed in an Appendix.

\section{The general helicity formula for $\gamma^* \gamma^*$
collisions}
\label{sec:general}

The general helicity formula for two-body or quasi two-body reactions
induced by the photon-photon interaction, i.e.
for processes of the type
$ e\,  e' \to e\, e'\, a \, b$
(where $ b$ may be a system of particles instead of a single one), as
shown by Fig. 1,
has been in the literature for a long time
\cite{carl,brown1,brown2,budnev2,carim1,carim2}. While at the
start this formula contains $3^4 = 81$ terms,
since the helicity matrix of either
photon is composed of $ 3 \times 3$ elements, that number is considerably
reduced by applying first principles,
namely hermiticity, parity conservation and
rotational invariance. Gathering together terms which have the same
behaviour with respect to azimuthal angles, we get a 13-term formula :
\begin{equation}
\frac{s Q^2 Q'^2}{e^8}\, d\sigma\, = \,
\frac{1}{(1-\epsilon)(1-\epsilon')}\,  F \, d\text{Lips}
\end{equation}
with :
\begin{eqnarray}
F & = & F_1 -2\sqrt{\epsilon(1+\epsilon)}F_2\cos{\varphi_a}
                -2\epsilon F_3\cos{2\varphi_a} \nonumber \\
  &\, & +2\sqrt{\epsilon'(1+\epsilon')}F_4\cos{(\varphi - \varphi_a)}
                -2\epsilon' F_5 \cos {2(\varphi -\varphi_a)}\nonumber\\
  & \, & - 2\sqrt{\epsilon(1+\epsilon)\epsilon'(1+\epsilon')}
         F_6 \, \cos \varphi
                        +\epsilon\epsilon' F_7 \cos 2\varphi \nonumber \\
  &\, & -2\sqrt{\epsilon(1+\epsilon)\epsilon'(1+\epsilon')}
                        F_8 \cos (2\varphi_a -\varphi)
        +\epsilon\epsilon' F_9 \cos 2(2\varphi_a-\varphi)\nonumber \\
  &\, & -2\epsilon\sqrt{\epsilon'(1+\epsilon')} F_{10}\cos
                          (\varphi +\varphi_a)
                          +2\epsilon'\sqrt{\epsilon(1+\epsilon)}F_{11}\cos
                          (2\varphi-\varphi_a)\nonumber\\
  &\, & -2\epsilon\sqrt{\epsilon'(1+\epsilon')} F_{12} \cos
                          (3\varphi_a -\varphi)
                          +2\epsilon'\sqrt{\epsilon(1+\epsilon)} F_{13} \cos
                          (2\varphi -3\varphi_a)
\end{eqnarray}

Here the quantities $F_i$ ($i = 1 ... 13$)
are linear combinations of elements of
the helicity tensor associated with the process
$\gamma \gamma' \to a \, b $ ;
these quantities,
which are typical structure functions
($F_1$ is the diagonal structure function, while all others may
be called ``interference structure functions''),
are given as follows :
\begin{eqnarray}
F_1 & = & F_{++,++} + F_{++,--} + 2\epsilon F_{00,++}
                         + 2 \epsilon' F_{++,00} + 2 \epsilon \epsilon'
F_{00,00}\, ;
                         \nonumber \\
F_2 & = & \Re  e (F_{+0,++} - F_{0-,++} + 2\epsilon'F_{+0,00})\, ; \qquad
F_3 = \Re e F_{+-,++} +\epsilon' F_{+-,00}\, ; \nonumber \\
F_4 & = & \Re e (F_{++,+0} -F_{++,0-}+2\epsilon F_{00,+0})\, ; \qquad
F_5 = \Re e F_{++,+-} +\epsilon F_{00,+-}\, ; \nonumber \\
F_6 & = & \Re e (F_{+0,+0} -F_{+0,0-})\, ; \qquad F_7 =  F_{+-,+-}\, ;\qquad
F_8 = \Re e (F_{+0,0+}-F_{+0,0-})\, ;\nonumber \\
F_9 & = & F_{+-,-+}\, ; \qquad F_{10} = \Re e F_{+-,+0}\, ; \qquad F_{11}
 = \Re e F_{+0,+-}\, ; \nonumber \\
 F_{12} & = & \Re e F_{+-,0+}\, ; \qquad F_{13} = \Re e F_{0+,+-}\, ;
\end{eqnarray}
where the tensor elements $ F_{m\bar{m},n\bar{n}}$  are defined as
\begin{equation}
F_{m\bar{m},n\bar{n}}= \sum {\cal M}_{mn}^{(\gamma \gamma' \to a b)}
\, {\cal M}_{\bar{m}\bar{n}}^{* (\, \gamma \gamma' \to a b)}
\end{equation}
calling $ m (\bar{m})$ resp. $n (\bar{n})$ the helicities of
$\gamma$  resp. $\gamma'$ ; the symbol  $\sum $
indicates summation over  the spin states of $ a$  and $ b$.
The elements
$F_{m\bar{m},n\bar{n}}$  depend on $ W^2, Q^2, Q'^2$ and $\chi$,
\mbox{defining :} $ W^2 = (q  + q')^2,\,  Q^2 = - q^2, \,
Q'^2 = - q'^2 $  (where $ q$  and $ q'$ are the respective
four-momenta of $\gamma$ and $ \gamma'$), and calling $\chi$
the polar angle of $ a$ with respect to $\gamma$
in the $\gamma \gamma'$  center-of-mass frame.
Let us remark that hermiticity entails the relation
$F_{m\bar{m},n\bar{n}} = F^*_{\bar{m}m,\bar{n}n}$, while
parity conservation and rotational invariance entail
$F_{m\bar{m},n\bar{n}} = (- 1)^{m+\bar{m}+n+\bar{n}}
F_{-m\, -\bar{m},-n\, -\bar{n}}$.

In addition we use the following notations : $\varphi$
is the azimuthal angle of $e'$ with respect to $e$ in the
$\gamma \gamma'$ c. m. frame with the $z$ axis oriented along the
three-momentum of the photon $\gamma$ ;
$\varphi_a $ is the azimuthal angle of $ a$
with respect to $e$ in the same frame.
The polarization parameters $\epsilon$ and $ \epsilon'$
of, respectively, the photons $\gamma$ and $\gamma'$
are given as follows :
\begin{eqnarray}
\epsilon & = & \frac{2x's(x's-W^2-Q^2-Q^{'2})+2Q^2Q^{'2}}
{(x's-W^2-Q^2-Q^{'2})^2 +x'^2s^2 - 2 Q^2Q^{'2}}
\nonumber \\
\epsilon' & = & \frac{2xs(xs-W^2-Q^2-Q^{'2})+2Q^2Q^{'2}}
{(xs-W^2-Q^2-Q^{'2})^2 +x^2s^2 - 2 Q^2Q^{'2}}
\end{eqnarray}
where $s$ is the total energy squared in the overall
center-of-mass frame, while $x$ and $x'$ are defined
as: $x=(q\cdot p'_0)/(p_0\cdot p'_0)$,
$x'=(p_0\cdot q')/(p_0\cdot p'_0)$,
calling $p_0, p'_0$ the respective four-momenta of $e_0, e'_0$.
Notice that in (5) we have neglected
the electron mass ; actually one has $\epsilon$ ($\epsilon') \to 0 $
when $Q^2$ ($Q^{'2} $) reaches its minimal value.

The Lorentz-invariant phase space is given by
\begin{equation}
d\text{Lips} = \frac{1}{(2\pi)^{3n_f-4}} \, \delta^4\,
\biggl(\sum_ip_i -\sum_f p_f\biggr)
\, \prod_f \frac{d^3{\bf p}_f}{p_f^0}
\end{equation}
where $n_f$ is the number of final particles, and
$ p_i, p_f $ are the respective four-momenta of
initial and final particles, while the superscript $0$
indicates the energy component. (If $b$ is a system of
particles, an integration over all corresponding 3-momenta,
except one, is implicitly included.)

Finally $ W^2 $ is given by
\begin{equation}
W^2 = - Q^2 -Q^{'2} + \frac{Q^2Q^{'2}}{s}
+xx's -  2QQ^{'}\sqrt{(1-x)(1-x')}
\cos \varphi^{\text{lab}}
\end{equation}
where $\varphi^{\text{lab}}$ is the azimuthal angle of $ e'$
with respect to $ e $ in the lab frame with the $z$ axis oriented
along the three-momentum of the incident electron $e_0$.

\section{Particular  configurations}
\label{sec:config}
\subsection{Configuration 1 :\protect\\
Double-tag measurement, extrapolating the central-detector
acceptance  to  4 $\pi$}

We here assume a measurement where both outgoing electrons are
tagged, and a so-called ``unfolding'' procedure is used in order to extrapolate
the acceptance of the central detector to $4\pi$.
This allows one to integrate formula (2) over $\varphi_a$
between $0$ and $2\pi$, so that one obtains the
three-term formula (see Ref.~\cite{balakin}) :
\begin{equation}
(2\pi)^{-1}\int F \, d\varphi_a =  F_1-2\sqrt{\epsilon(1+\epsilon)
\epsilon'(1+\epsilon')}\, F_6 \, \cos \varphi
+\epsilon \epsilon' \, F_7 \, \cos 2\varphi
\end{equation}

Actually formula (8) can hardly be exploited in that form for an
azimuthal-correlation study, for the following reason :
Since $F_6$ involves longitudinal
(helicity 0) components of both photons, it
can be shown by general arguments (see section III of Ref.~\cite{carim2})
to stay
non-negligible only when both outgoing electrons are tagged at
large angles (i.e. $ Q, Q^{'} \approx  W$)~\cite{note}.
But in that case $W$ depends on $\varphi$
(as can be inferred from formula (7), since $\varphi^{\text{lab}}$
is correlated with $\varphi$), and therefore the functions  $F_i$,
as well as their coefficients, also depend on $\varphi$ .
It is then obvious that
it would become much more complicated to extract the structure
functions from the analysis of the $\varphi$ distribution
(all the more as the integrated cross sections are expected to
be very small in that kinematic situation).

One is thus led to consider the case where
at least one of the outgoing electrons
(for instance $ e'$) is tagged at small angle (i.e. $ Q^{'}\ll W/2$).
We now use formula (7) and in addition the relation
\begin{equation}
\cos \varphi = \cos \varphi^{\text{lab}}+\frac{QQ^{'}}{W^2+Q^2}
\frac{2 -x -x'}{\sqrt{(1-x)(1-x')}}\sin^2\varphi^{\text{lab}}
+O\biggl(\frac{Q^{'2}}{W^2}\biggr)
\end{equation}
That relation shows that - except for a small
range near $\varphi^{\text{lab}} = \pi/2$,
and for marginal ranges of  $x$ and $x'$ that one can
suppress by setting upper limits on those variables -
one is allowed to identify $\varphi$ with $\varphi^{\text{lab}}$.
We thus get :
\begin{equation}
W^2 = -Q^2 +xx's - 2Q Q^{'} \sqrt{(1-x)(1-x')} \, \cos \varphi
+O (Q^{'2})
\end{equation}

Still one may consider two different experimental situations :
(A) both electrons are tagged at small angle $(Q, Q^{'}\ll W/2)$ ;
(B) $ e'$  is tagged at small angle, $ e$ at large angle
($Q^{'}\ll W/2,  Q \lesssim W$). In both cases $W$ tends to become
independent of  $\varphi$, i.e. :
\begin{equation}
\text{(A)}\,\, W^2 \simeq x x' s \qquad  \text{resp.}\qquad
\text{(B)}\,  \, W^2 \simeq -Q^2 +x x' s,
\end{equation}
while at the same time the condition for neglect of
longitudinal contributions (Ref.~\cite{carim2}, formula (3.5)),
i.e. $  Q^{'}\ll (W^2 + Q^2 - Q^{'2})/(2 W)$,  is satisfied, so that
$F_6$  goes to zero. One is then left with the two-term formula
\begin{equation}
(2\pi)^{-1}\int F \, d\varphi_a =
 F_1 +\epsilon \epsilon' F_7 \cos 2\varphi
\end{equation}
where in case (A) one has
\begin{equation}
F_1 = F_{++,++} +F_{++,--}\, ; \qquad F_7 = F_{+-,+-}\, ;
\end{equation}
with the helicity-tensor elements depending only on $W^2$ and $\chi$,
while in case (B) one gets
\begin{equation}
F_1 = F_{++,++} +F_{++,--}+2\epsilon F_{00,++}\, ;
\qquad F_7 = F_{+-,+-}\, ;
\end{equation}
with the helicity-tensor elements depending on $W^2$, $Q^2$  and  $\chi$.

$F_1$, being composed only of diagonal elements of the
helicity-tensor, is to be identified, apart from kinematic factors,
with the differential cross section
$d\sigma_{\gamma\gamma}/d(\cos \chi )$  in case (A),
resp.  $(d\sigma_T + \epsilon d\sigma_L)/d(\cos \chi )$
in case (B), for the reaction $\gamma \gamma' \to a b $ ;
for instance, in the case where ($a,b$) is a pair of charged particles,
one gets
\begin{equation}
\frac{d\sigma_{\gamma\gamma}}{d(\cos\chi)}\quad \hbox{\text{resp}}\quad
\frac{d\sigma_T +\epsilon d\sigma_L }{d(\cos\chi)}
\quad = \quad \frac{\Lambda^{1/2}(W^2,m_a^2,m_b^2)}{64\pi W^4} \, F_1
\end{equation}
where $\Lambda$ is defined as :
$\Lambda(A,B,C) = A^2 +B^2 +C^2 -2AB -2AC -2BC $.
That cross section can obviously be determined as well in
the no-tag resp. the single-tag mode.
Double tagging, with the constraints defined in order to allow
for an azimuthal-correlation analysis, provides the
possibility of extracting an additional structure
function ($F_7$) ; the price to pay is, of course, a lowering of
the yield obtained.

Notice that, at $Q^{'} \ll W/2 $, formula (5) becomes
considerably simplified,
as one gets
\begin{equation}
\epsilon = \frac{1-x}{1-x+x^2/2} ;
\qquad \epsilon' = \frac{1 -x'}{1-x'+x'^2/2}
\end{equation}
Finally let us remark that, since
$\varphi \simeq \varphi^{\text{lab}}$, as shown by formula (9), and
since one may assume the electron-tagging systems to be
cylindrically symmetric, there should be no distorsion,
due to the apparatus,  of the $\varphi$  distribution.

\subsection{Configuration 2 :\protect\\  Single-tag  measurement}

In that configuration the untagged (or antitagged) electron, e.g. $e'$, is
predominantly emitted very close to $0^{\circ}$, so that
$Q^{'} $ becomes essentially negligible as compared with $W/2$,
which has the obvious consequences
that :  (i)  $W$  becomes independent of
$\varphi$ (see again formula (10)) ;
(ii)  $\varphi \simeq \varphi^{\text{lab}}$ (formula (9))
and therefore, assuming here again the electron tagging
system to be cylindrically symmetric, there should be no
distorsion of the $\varphi$  distribution due to the apparatus.
We may thus integrate formula (2) over $\varphi$  between $0$ and $2\pi$,
so that we obtain the three-term formula :
\begin{equation}
(2\pi)^{-1}\int F \, d\varphi = F_1-2\sqrt{\epsilon(1+\epsilon)} F_2
\cos \varphi_a - 2\epsilon F_3 \cos 2\varphi_a
\end{equation}
with
\begin{eqnarray}
F_1 & = & F_{++,++} +F_{++,--} + 2 \epsilon F_{00,++}\, ;
\qquad F_2 = \Re e (F_{+0,++} - F_{0-,++})\, ; \nonumber \\
F_3 & = & \Re e F_{+-,++}\, .
\end{eqnarray}
Again  $\epsilon, \epsilon' $  are given by formula (16).
In Ref.~\cite{ong1} formula (18) has been applied to muon and pion pair
production, using two different models for the latter.
However, in that paper, the problem of experimental constraints
to be applied in order to suppress the distorsion of the $\varphi_a$
distribution, induced by  the limited acceptance of the
central detector, was left aside.
Actually one should here assume that the latter
has an ``almost $4 \pi$'' acceptance.
Yet a kinematic study shows that, even if the acceptance
cuts of the central detector remain very small, the cuts induced by
them in the polar emission angle of the particles produced in
the $\gamma \gamma'$ c. m.
frame may be large, not confined to the margins of phase space,
and azimuth-dependent (as was already noticed in Ref.~\cite{diber}).
Therefore additional constraints should be imposed in order to
minimize the latter cuts. This problem, which appears as
well in configuration 4, is treated in the Appendix of this paper.

Notice that in the single-tag case $W$ can only be determined
by measuring all particles produced. When this is not possible
(in the case of multi-particle final-states), the fact that
$W_{\rm{vis}} \neq W $ is an additional source of
complications.

Let us mention that a measurement of this type, involving muon pair
production, has recently been performed by the L3 Collaboration
at LEP (CERN) \cite{leonardi}.

\subsection{Configuration 3 :\protect\\
Double-tag measurement at small angles}

We here assume that the electron-tagging angles are small enough to
ensure that\\ \noindent  $Q, Q^{'}\ll W/2$, so that in formula (2)
we may neglect all longitudinal helicity-tensor elements,
i.e. those with at least one 0 subscript.
We are thus left with the
five-term formula (see Ref.~\cite{budnev2}, Eq. (5.33)) :
\newpage
\begin{eqnarray}
F  =  F_1 & - & 2 \epsilon F_3 \cos 2\varphi_a - 2 \epsilon' F_5
                  \cos 2(\varphi -\varphi_a)\nonumber \\
  & + & \epsilon \epsilon' F_7 \cos 2\varphi +\epsilon \epsilon'
                  F_9 \cos 2(2\varphi_a-\varphi)
\end{eqnarray}
with
\begin{eqnarray}
F_1 & = & F_{++,++} + F_{++,--}\, ; \qquad F_3 = \Re e F_{+-,++}\, ;
\nonumber \\
F_5 & = & \Re e F_{++,+-}\, ; \qquad F_7 = F_{+-,+-}\, ;
\qquad F_9 = F_{+-,-+}\, ;
\end{eqnarray}
where the helicity-tensor elements depend on $W^2$ and $\chi$ ;
$ \epsilon$, $\epsilon'$ are again given by formula (16).
At this point we note that, if the whole range
$0 < \varphi < 2\pi $, $ 0 < \varphi_a < 2\pi $ is available,
one may use the orthogonality of the functions
\{$1, \cos 2\varphi_a , \cos2(\varphi-\varphi_a), \cos 2\varphi ,
\cos 2(2\varphi_a -\varphi )$\} in that range and thus
extract $F_1 ..... F_5 $ by projecting formula (19)
on those functions. However it seems preferable, in order
to get more information, to apply what we call
``azimuthal selection'', i.e. to select individual
distributions with respect to the azimuthal angles
involved, as follows :
let us assume, for instance, that we are interested in the
distribution with respect to the azimuthal angle
$\hat{\varphi} = 2 \varphi_a - \varphi$.
Switching from our system of azimuthal variables $\varphi, \varphi_a$
to the system $\varphi , \hat{\varphi}$, we write :
\begin{eqnarray}
F  =  F_1 & - & 2 \epsilon F_3 \cos (\varphi+\hat{\varphi})
                - 2 \epsilon' F_5
                  \cos (\varphi -\hat{\varphi})\nonumber \\
  & \, & \qquad  +\, \epsilon \epsilon' F_7 \cos 2\varphi +
  \, \epsilon \epsilon' F_9 \cos 2\hat{\varphi}
\end{eqnarray}

Integrating over $\varphi$ between $0$ and $2\pi$, we then get :
\begin{equation}
(2\pi)^{-1}\int F \, d\varphi =  F_1 +\epsilon \epsilon'
F_9 \cos 2 \hat{\varphi}
\end{equation}

Similarly we can select three other azimuthal-angle distributions :
\begin{eqnarray}
& &(\text{using the system of variables}\, \varphi, \varphi_a) :
\nonumber \\ & &
\qquad (2\pi)^{-1}\int F d \varphi =  F_1 - 2\epsilon
F_3 \cos 2\varphi_a \\
& &(\text{using the system of variables}\, \varphi,
\varphi_a'=\varphi -\varphi_a) :
\nonumber \\ & & \qquad (2\pi)^{-1}\int F d\varphi =
F_1 -2 \epsilon' F_5 \cos 2\varphi_a' \\
& &(\text{using the system of variables}\, \varphi,\varphi_a) :
\nonumber \\ & &
\qquad (2\pi)^{-1}\int F d\varphi_a =  F_1 +\epsilon\epsilon'
F_7 \cos 2\varphi
\end{eqnarray}

Here we have again made use of the fact that the $\varphi$-dependence
of $W$ can be neglected according to formula (10).
In addition we have implicitly assumed that there
is no distorsion of the  $\varphi $  distribution due to the
apparatus (i.e. $\varphi$ varies
between $0$ and $2\pi $ whatever the values of the other
variables within the phase space considered) ;
that assumption is justified, here again, by the
fact that  $ \varphi \simeq \varphi^{\text{lab}}$
(formula (9)), and that we may suppose both electrons
to be tagged in a cylindrically symmetric way.

Finally we have implicitly assumed that there is no distorsion of the
$\varphi_a$
distribution. This assumption, as already discussed in Ref.~\cite{ong2},
requires a somewhat more stringent condition to be imposed on
$Q$ and $Q^{'}$, namely : $ Q, Q^{'}\ll (W/2)\sin \chi$ ;
that means that the transverse momenta of both photons in
the lab frame can be neglected with respect to the transverse
momentum of $a$ in the $\gamma \gamma'$ c. m. frame.
Under that condition
the $ \gamma \gamma' $  collision axis tends to be identical
with the {colliding-beam} axis ; consequently
one gets
$\varphi_a \simeq \varphi_a^{\text{lab}}$,
and it thus becomes sufficient to assume that the central
detector is, as well, cylindrically symmetric. The condition here
defined implies of course that an appropriate lower limit is assigned
to $\sin\chi$.

As one sees from formulas (21)-(24), azimuthal selection should allow
one, in the  configu- ration considered, to extract four azimuthal-angle
distributions, and correspondingly five structure functions, from the data
obtained in a single measurement. It must however be remarked
that because of left-right symmetry the  $\varphi_a$  and  $\varphi_a'$
distributions, and correspondingly  the
structure functions  $F_3$  and $F_5$, should be identical.
Thus, we get in fact three independent azimuthal correlations,
allowing for the determination of four independent structure functions.

In Ref.~\cite{ong2} formulas (22)-(25) have been applied to lepton
and pion pair production.

It is to be mentioned that an experiment of that type is presently being
prepared at the low-energy electron-positron collider
DA$\Phi$NE at Frascati \cite{bellucci}.

\newpage
\subsection{Configuration 4 :\protect\\
Double-tag  measurement with  one  electron  tagged at
large  angle  and  the other one at small angle}

As in case (B) of configuration 1, we here assume that
the electron $e$ is tagged at large angle ($Q \lesssim W$),
while $e'$  is assumed to
be tagged at small angle ($Q^{'} \ll   W/2$). Thus $W$ becomes
practically indendent of $\varphi $, according to formula
(10), while at the same time all helicity-tensor
elements with at least one $0$ helicity subscript for the
right-hand photon tend to vanish. Formula (2) is then reduced
to an eight-term formula :
\begin{eqnarray}
F & = & F_1  - 2\sqrt{\epsilon(1+\epsilon)} F_2 \cos \varphi_a
             -2 \epsilon F_3 \cos 2\varphi_a \nonumber \\
&\, & - 2\epsilon' F_5 \cos 2(\varphi-\varphi_a) +
                \epsilon\epsilon' F_7 \cos 2\varphi +
                \epsilon \epsilon' F_9 \cos 2(2\varphi_a - \varphi)\nonumber \\
&\, & +2 \epsilon'\sqrt{\epsilon(1+\epsilon)} F_{11}
\cos (2 \varphi -\varphi_a) + 2 \epsilon' \sqrt{\epsilon(1+\epsilon)}
F_{13} \cos (2\varphi -3 \varphi_a)
\end{eqnarray}
with
\begin{eqnarray}
F_1 & = & F_{++,++} +F_{++,--} + 2\epsilon F_{00,++}\, ; \qquad
                         F_2= \Re e (F_{+0,++} - F_{-0,++})\nonumber \\
F_3 & = & \Re e F_{+-,++}\, ;
\qquad F_5 = \Re e F_{++,+-} + \epsilon F_{00,+-}\, ;
\qquad F_7 = F_{+-,+-}\, ;\nonumber \\
F_9 & = & F_{+-,-+}\, ; \qquad F_{11} = \Re e F_{+0,+-}\, ;
                         \qquad F_{13} = \Re e F_{0+,+-}\, ;
\end{eqnarray}

\noindent where the helicity-tensor elements depend on
$W^2, Q^2$ and $\chi$ ;
$\epsilon, \epsilon'$ are again given by  formula (16).

Azimuthal selection, applied in the same way as in III.C, makes it
possible to derive from formula (26) six independent
azimuthal correlations, from which eight independent
structure functions can be extracted. Those correlations
are the following :
\begin{eqnarray}
& & (\text{using the system of variables}\, \varphi, \varphi_a) :
\nonumber \\ & &
\qquad (2\pi)^{-1}\int F d\varphi =  F_1 -2\sqrt{\epsilon(1+\epsilon)}
F_2 \cos \varphi_a - 2 \epsilon F_3 \cos 2\varphi_a \\
& & (\text{using the system of variables}\,  \varphi, \varphi -\varphi_a) :
\nonumber \\ & &
\qquad (2\pi)^{-1} \int F d\varphi =  F_1 -2\epsilon'
F_5 \cos 2(\varphi - \varphi_a)\\
& & (\text{using the system of variables}\, \varphi,  \varphi_a ) :
\nonumber \\ & &
\qquad
(2\pi)^{-1} \int F d \varphi_a =  F_1 +\epsilon \epsilon' F_7 \cos
2\varphi\\
& & (\text{using the system of variables}\, \varphi, 2\varphi_a -\varphi) :
\nonumber \\ & &
\qquad (2\pi)^{-1} \int F d\varphi =  F_1 +\epsilon \epsilon'
F_9 \cos (2\varphi_a - \varphi)\\
& & (\text{using the system of variables}\, \varphi, 2\varphi -\varphi_a) :
\nonumber \\ &  &
\qquad (2\pi)^{-1}\int F d \varphi = F_1
+2\epsilon'\sqrt{\epsilon(1+\epsilon)} F_{11} \cos (2\varphi -
\varphi_a) \\
& & (\text{using the system of variables}\, \varphi, 2\varphi -3\varphi_a) :
\nonumber \\ & &
\qquad (2\pi)^{-1}\int F d \varphi = F_1 +2\epsilon'
\sqrt{\epsilon(1+\epsilon)} F_{13} \cos (2\varphi - 3\varphi_a)
\end{eqnarray}

Here again we have assumed that there is no distorsion of the
$\varphi$-distribution due to the apparatus, since
$\varphi \simeq \varphi^{\text{lab}}$ (formula (9))
and we may consider that
both electrons are tagged in a cylindrically symmetric way.

On the other hand (as in configuration 2),
given the limited acceptance of the central detector,
ensuring the absence of
distorsion of the $\varphi_a$ distribution
is a critical problem that will be treated in the Appendix.

\section{Applications  of   configuration  4}

        We shall now consider practical applications of the 8-term
formula (26) computed in III.D.  The processes considered are, here
again, $\gamma \gamma$ production of muon and pion pairs.
In that case one gets from formula (6),
assuming $m_a^2, m_b^2 \ll W^2 $ :
\begin{equation}
d\text{Lips} = \frac{1}{(4\pi)^7}
\, dx\, dx' \, dQ^2 \, dQ^{'2}\, d\varphi^{\text{lab}}\,
d\cos\chi d\varphi_a
\end{equation}

        Equating $\varphi^{\text{lab}}$ with $\varphi $ (see (9)), and
taking account of (10) (with $Q^{'} \to 0 $),
this expression transforms into :
\begin{equation}
d\text{Lips} = \frac{1}{(4\pi)^7} \frac{dx}{x}\frac{dW^2}{s}
dQ^2 \, dQ^{'2} \, d\varphi d\cos \chi \, d \varphi_a
\end{equation}

Returning  to formula (1), and using (16) and again (10), we obtain
\begin{eqnarray}
\sigma
= & & \frac{\alpha^4}{2\pi^3}
\int W\, dW \, \frac{dQ}{Q}\frac{dQ^{'}}{Q^{'}}\frac{dx}{x}\, d\cos\chi
\nonumber \\
& & \cdot
\frac{1}{(W^2+Q^2)^2} \biggl(1 - x +\frac{x^2}{2}\biggr)
\biggl(1-\frac{W^2+Q^2}{xs}
+\frac{(W^2+Q^2)^2}{2x^2s^2}\biggr)\, F
\, {d\varphi\,  d\varphi_a}
\end{eqnarray}

For our computations we have fixed a number of limits on the
various integration parameters chosen. Those limits are the following :
\begin{itemize}
\item[(i)]
$Q^{'} > 5\times 10^{-3} E_0$
(corresponding approximately to a minimal tagging angle
of $5$ mrad for $ e'$), and  $Q^{'} <  W/20$
(in order to justify the neglect of $\varphi$-dependence
of  $W$ in formula (10), and at the same time to motivate the
neglect of longitudinal polarization of the photon  $\gamma'$).

\item[(ii)]
$Q <  W$ (again because of formula (10)),  and $Q > W/4$
(in order to ensure a significant contribution of the
longitudinal polarization of $\gamma$).

\item[(iii)]
$W^2 > 4 m E_0$ (see Appendix).

\item[(iv)]
Once again because of formula (10), we set :  $x < 0.7$ ; $ x' < 0.7$,
or equivalently\\ $x > (W^2 + Q^2)/(2.8 E_0^2)$.

\item[(v)]
$x > Q/2 E_0$  and $x < (W^2 + Q^2)/(4 Q E_0)$  (see Appendix) ;
these limits, combined with those defined in (iv), also induce
further limitations of $Q$ and $W$.

\item[(vi)]
Finally, because of the particular sensitivity of polar-angle ranges
close to $0$ resp. $\pi$ with regard to possible cuts located there,
we set : $ - 0.9 < \cos \chi < 0.9$.
\end{itemize}

Using formula (36) and replacing  $ \int F d\varphi $, resp.
$\int  F d\varphi_a $, by their expressions given in formulas
(28)-(33), and then integrating over the variables $\chi, x, Q', Q$
and  $W$, we have obtained, for each particular process and
model considered, the six angular distributions  $d\sigma/d\Phi$,
where $\Phi$ is to be identified respectively with
$\varphi_a, \varphi-\varphi_a, \varphi, 2\varphi_a -\varphi,
2\varphi -\varphi_a, 2\varphi - 3\varphi_a $.
In Figures 2-5 we are showing, for  $\Phi$ ranging
between $ 0$ and  $\pi$
(notice that, in the $\Phi $ range between  $\pi$  and  $2\pi$,  symmetric
values are obtained), the normalized distributions
$d\hat{\sigma}/d\Phi$, defined in such a
way that  $\int (d\hat{\sigma}/d\Phi) d\Phi = \pi $.

Let us remark that, since the structure functions $F_2, F_{11}$
and $F_{13}$ are changing sign with $\cos\chi$, the distributions
in $\varphi_a, 2\varphi -\varphi_a, 2\varphi -3\varphi_a$ have
been computed by integrating only over half of the $\cos\chi$
range, i.e. $0 < \cos\chi < 0.9 $.

Fig. 2 shows the azimuthal correlations obtained for muon pair
production, assuming a beam energy of $ 5 $\  GeV. We have checked that the
shapes of those curves remain practically unchanged when one goes over to a
much higher beam energy ($E_0 = 100$ GeV).

Figs. 3 and 4 show, again for $E_0 = 5 $ GeV, the
corresponding correlations
obtained for  $\pi^+ \pi^-$  production, as computed respectively in two
different models : the perturbative-QCD model proposed by Brodsky and
Lepage \cite{brodsky}, as extended by Gunion et al. \cite{gunion}
for  $\gamma \gamma^* $ collisions, with the pion wave function
given by Chernyak and Zhitnitsky \cite{chernyak} ;
and the finite-size model of Poppe \cite{poppe},
where the amplitude of the Born-term calculation
is multiplied by an overall form factor.
Notice that here the lower limit introduced for  $W$  is
$4 m_{\pi} E_0 \simeq 1.7$ GeV.

Fig. 5 shows the analogous curves obtained for  $\pi^0 \pi^0$
production at the
same beam energy, using again the Brodsky-Lepage model with the
Chernyak-Zhitnitsky wave function.

The values of integrated cross sections (obtained by replacing $F$
by $F_1$ in (36)) are given in Table 1, where they
are compared  with those of the ``theoretical background'' (see
Ref.~\cite{parisi}), i.e. of the contribution of the Feynman diagram
shown in Fig. 6, computed
within the same kinematic limits. For $\pi^+ \pi^- $
production, we here use a simple
VDM model involving only $\rho$-exchange.
It is seen that this background remains
insignificant as compared to the signal (in the case of
$\pi^0\pi^0$ production, it is of
course strictly zero). As regards the interference
term between the diagrams of Figs. 1 and 6, it is reduced
to zero if, instead of identifying particle  $a$  with the
muon (resp. pion) of either positive or negative charge, we
average over those two options.

Notice that in the figures of Table 1 we have included a factor of 2 in
order to account for the possibility of performing a symmetric
measurement ($e$  tagged at large angle, $ e'$
at small angle ; and conversely).

Finally it is to be emphasized that, if one uses a Monte-Carlo
program taking account of all experimental acceptances, one would
certainly be allowed to suppress some of the severe kinematic
restrictions here introduced.

\section{Discussion  and  conclusion}

        Azimuthal correlations are, to a large extent, a new approach to the
study of photon-photon collisions. Till now practically all experimental
efforts in two-photon physics have been concentrated on measuring
$\gamma \gamma $ cross
sections or the ``photon structure function''  $F_2^{\gamma}$
(i.e. the  $\gamma \gamma^*$ total hadronic
cross section), which correspond to the diagonal structure
function $F_1$ in the
formalism presented here. Azimuthal correlations should allow one to
determine, depending on the configuration considered, one, two, three or
seven additional independent structure functions. It should be
emphasized that in principle
$F_1$ does not contain a larger amount of physical information than
the others. Its special status arises only from the fact that
it is more easily determined ; in addition,
due to the Schwarz inequality, it is larger than (or at least equal to)
any of the interference structure functions.
In that sense it appears as the {\it primus inter
pares}, but not more. In other words :
azimuthal correlations should allow one to multiply the
physical information, obtained in a number of  two-photon processes,
by a factor of 2, 3, 4 or 8.

        As can be seen for instance by comparing Figs. 3 and 4, two different
dynamic models for a given process can lead to very different shapes of the
azimuthal-correlation curves, implying that the values of the structure
functions involved are very different as well. Thus azimuthal correlations
should provide one with a powerful tool for checking dynamic models.
Actually we may safely state that no model will survive this kind of check
if it is not entirely realistic.
        There is however a price to pay for this achievement : there are more
or less stringent experimental constraints that should be satisfied. In
configurations 1 and 3, what is required is the possibility of tagging the
outgoing electrons (and of measuring their azimuthal angles) at small
scattering angles. In configuration 2, an
``almost $4\pi$'' central detector appears necessary.
In configuration 4 both requirements are combined, and
in addition, since the measured cross section is sharply reduced by
significant cuts in most of the parameters to be measured, a very high
machine luminosity is required.

However, as already noticed above, the use of a Monte-Carlo program
should allow one to relax to some extent the constraints here defined
and thus to increase the integrated cross section.

In any case, and whatever the particular $\gamma \gamma $ process and
the configuration considered, azimuthal-correlation measurements
in muon pair production under the same conditions should always
be used as a test, so as to check the validity of the approximations
applied. Notice that in general it would be more problematic
to use electron
pairs for that purpose, given the
complications due to exchange between scattered and produced
electrons.

Let us finally remark that, for checking perturbative
QCD, it appears particularly interesting to look for azimuthal
correlations in the two-photon production of quark pairs
(notice that, in the quark-parton model, the azimuthal-correlation
curves predicted for the production of light-quark pairs are the same
as for $\mu^+ \mu^- $ production).
Measuring azimuthal angles of jets is probably a difficult, but not
impossible, task for experimentalists. Another option would
be to investigate azimuthal correlations in the inclusive production of
one hadron (plus anything).

\acknowledgements

The authors are grateful to Dr. A.~Courau for illuminating
discussions and precious advice.
One of us (O.~P.) would like to thank the theory group of
the ``Laboratoire de Physique Corpusculaire, Coll\`ege de France''
for the warm hospitality extended to him.

\newpage
\appendix
\section*{
Experimental  constraints  to  be  applied  to  configurations  2
and  4}

Our problem is how to minimize azimuth-dependent cuts in $\cos \chi$ ,
induced by the limited acceptance of the central detector.
For simplicity, we shall stick to the case of production of
particle-antiparticle pairs ($b = \bar{a}$).
Then we can use formulas (A10) - (A12) of the
Appendix of Ref.~\cite{carim2}.
We shall however rewrite those formulas, using slightly
different notations (and in addition, as regards formula (A12),
exchanging the variables pertaining to the left- and
right-hand vertex of Fig. 1). We thus get

\begin{eqnarray}
\beta \, \Lambda^{1/2}(W^2,-Q^2,-Q^{'2})
\, \cos{\chi} \,
& = &
Q^{'2}-W^2-Q^2
+4E_a\bigl\{E_0(1+\beta_a\cos\psi) - \nonumber \\
& & E^{'}\bigl[1 + \beta_a\cos\theta'\cos \psi
-\beta_a\sin\theta'\sin\psi
\cos (\varphi^{\text{lab}} -\varphi_a^{\text{lab}})\bigr]\bigl\}
\end{eqnarray}

\noindent with $E_a$, the lab energy of particle $a$, given by
\begin{equation}
E_a = \frac{E_X W^2 \pm \tilde{p}_X \bigg[W^4
-4m_a^2(E_X^2 -\tilde{p}_X^2)\biggr]^{1/2}}{2(E_X^2 -\tilde{p}_X^2)}
\end{equation}
where we have defined
\begin{eqnarray}
E_X & = & 2E_0 - E -E^{'} \\
\tilde{p}_X & = & - E( \cos\theta \cos \psi +\sin\theta \sin \psi
\cos\varphi_a^{\text{lab}})+ \nonumber \\
& & E^{'}\bigl[\cos\theta'\cos\psi -\sin\theta'\sin\psi\cos
(\varphi^{\text{lab}}-\varphi_a^{\text{lab}})\bigr]
\end{eqnarray}
and where in addition we have used the following definitions :
$\theta,\theta'$
are the lab scattering angles of $ e, e'$, while $\psi$ is the
lab polar emission angle of particle $a$ with respect to $e$ ;
$\beta = ( 1 - 4 m_a^2/W^2)$ ;
$\beta_a = (1 - m_a^2/E_a^2)^{1/2}$.
Notice that in formula (A2) only solution + is valid when
$W^2 > 2 m_a E_X$.

It is to be remarked that
$\cos \chi$  is  implicitly depending on
$\varphi , \varphi_a $
through its explicit dependence on
$\varphi^{\text{lab}}$ and $ \varphi_a^{\text{lab}}$
(see formulas (A9),
(A13) of Ref.~\cite{carim2}, showing the relations between
azimuthal angles in the lab and the c. m. frame).

In the configuration (2 or 4) considered here, we let $ Q^{'}$
(and correspondingly  $\theta'$) go to zero, so that formula (A1)
becomes
\begin{equation}
1 +\cos \chi \simeq \frac{4E_ax'E_0(1+\beta_a\cos\psi)}
{W^2 +Q^2}\simeq \frac{E_a(1+\beta_a\cos\psi)}{xE_0}
\end{equation}

On the other hand we assume $W^2 > 4 m_a E_0$, which entails
(according to (A3), setting
$x, x' > 0.7$) that $W^4 > 8 m_a^2 E_X^2$.
Therefore only solution + is to be considered in (A2) ;
in addition it becomes possible to make a convergent series
expansion, in $m_a^2/W^2$, of the radical in (A2).
That formula thus becomes
\begin{equation}
E_a \simeq \frac{W^2}{2(E_X -\tilde{p}_X)} - \tilde{p}_X\frac{m_a^2}
{W^2} + O(\frac{m_a^4}{W^4})
\end{equation}
where we shall retain only the first term on the right-hand side.
Substituting that expression of  $E_a$  into (A5), we get,
taking account of (A3), (A4) with $\theta' \to 0$ :
\vspace{0.1cm}

\begin{equation}
1 + \cos \chi = \frac{(1+\beta_a\cos \psi) W^2}
{2xE_0[2E_0-E(1-\cos\theta\cos\psi+\sin\theta\sin\psi
\cos\varphi_a^{\text{lab}})-E^{'}(1+\cos\psi)]}
\end{equation}

We now assume that the acceptance of the central detector
is given by : $ \psi_0 < \psi <  \pi  - \psi_0 $, with
$\psi_0 \ll $ 1 rad. We then compute the cuts in $\cos \chi $
induced by the forward and backward cut in $\psi$,
neglecting terms in $\psi_0^2$.

(i) Forward cut :
\begin{eqnarray}
\vert \Delta (\cos \chi)\vert &=& \bigg\vert (1+\cos \chi )_{\psi=\psi_0}
-(1+\cos\chi)_{\psi=0}\bigg\vert \nonumber\\
& & \nonumber \\
&\simeq & (1+\beta_a)\bigg\vert \frac{W^2}{2xE_0(2x'E_0 -Q^2/2E_0
+\psi_0E\sin\theta\cos\varphi_a^{\text{lab}})}-\frac{W^2}
{2xE_0(2x'E_0 -Q^2/2E_0)}\bigg\vert \nonumber\\
& & \nonumber \\
&\simeq& (1+\beta_a)\frac{W^2}{W^2+(1-x)Q^2}\,
\frac{\psi_0E\sin\theta\vert\cos\varphi_a^{\text{lab}}\vert}
{2x'E_0(1-Q^2/s)} < \frac{Q\psi_0}{x'E_0(1-Q^2/s)}
\end{eqnarray}

\vspace{0.3cm}
Setting  $x'> Q/[E_0(1-Q^2/s)]$, i.e.
$x < (W^2+Q^2)(1-Q^2/s)/(4QE_0)$, one is led to
$\vert \Delta \cos \chi \vert < \psi_0 $.
For simplicity (and since values of $Q^2/s$ much smaller than $1$ are
largely predominating), we
shall approximate the above-defined upper
limit of $ x$ by requiring only  $ x < (W^2 + Q^2)/(4 Q E_0)$.

\newpage
(ii) Backward cut :
\begin{eqnarray}
\vert \Delta (\cos \chi)\vert &=& \bigg\vert (1+\cos \chi )_{\psi
= \pi - \psi_0}
-(1+\cos\chi)_{\psi=\pi}\bigg\vert \nonumber\\
& & \nonumber \\
&\simeq & (1-\beta_a)\bigg\vert \frac{W^2}{2xE_0(2xE_0
+\psi_0E\sin\theta\cos\varphi_a^{\text{lab}})}-\frac{W^2}
{4x^2E_0^2}\bigg\vert \nonumber\\
& & \nonumber \\
&\simeq& (1-\beta_a)\frac{W^2}{4x^2E_0^2}\,
\frac{\psi_0E\sin\theta\vert\cos\varphi_a^{\text{lab}}\vert}
{2xE_0} < \frac{W^2Q\psi_0}{8x^3E_0^3}
\end{eqnarray}

\vspace{0.3cm}
Setting $x > \sup \, [W/(2 E_0), Q/(2 E_0)]$,
we are led here again to
$\vert \Delta \cos \chi \vert < \psi_0 $.
Notice that in configuration 4, having set  $Q \lesssim W$,
that condition simply becomes  $x > W/(2E_0)$.

It is easily checked that the lower and upper limit thus fixed
for $x$ are compatible without imposing any further constraint
on $W$ and $Q$.

With $\psi_0  = 0.1 $ rad, the azimuth-dependent cut in $\cos \chi$,
induced either by the forward or the backward cut in $x$,
is reduced to less than 5 \% of phase space.

We have verified a posteriori that, with the constraints here
defined, terms that we have neglected in the expressions of
$\cos \chi$ and $E_a$ have only negligible effects
on $\Delta \cos \chi$.

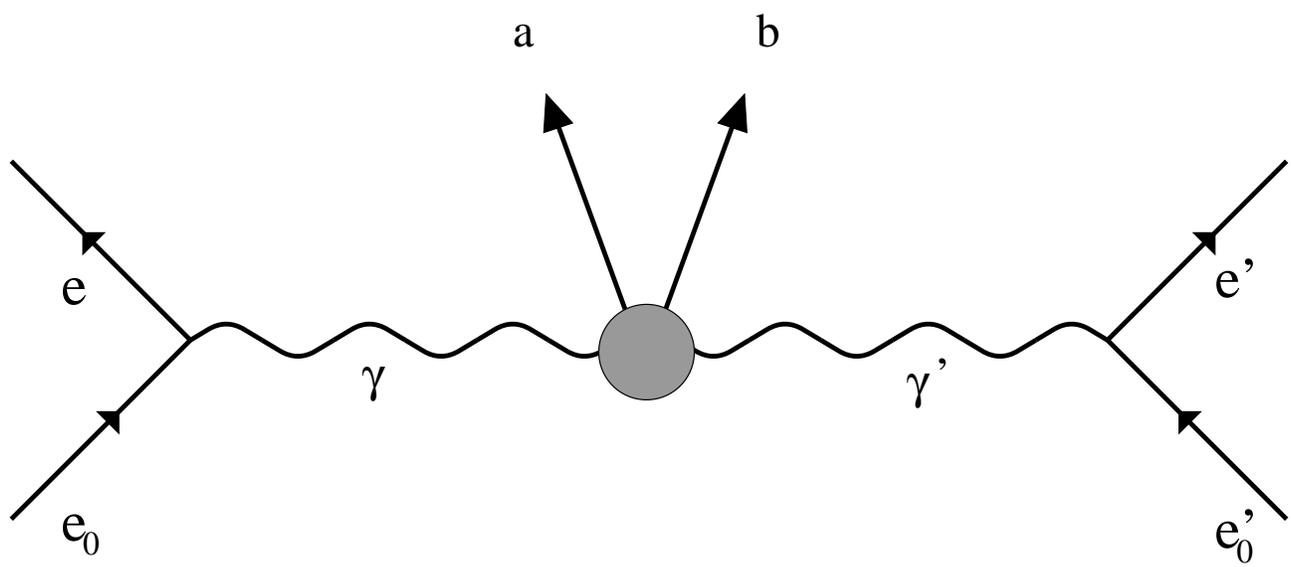
\begin{figure}
\caption{
Feynman diagram for the reaction
$ e \, e' \to e \, e' \, a \, b $
involving the exchange of two spacelike photons.
}
\end{figure}
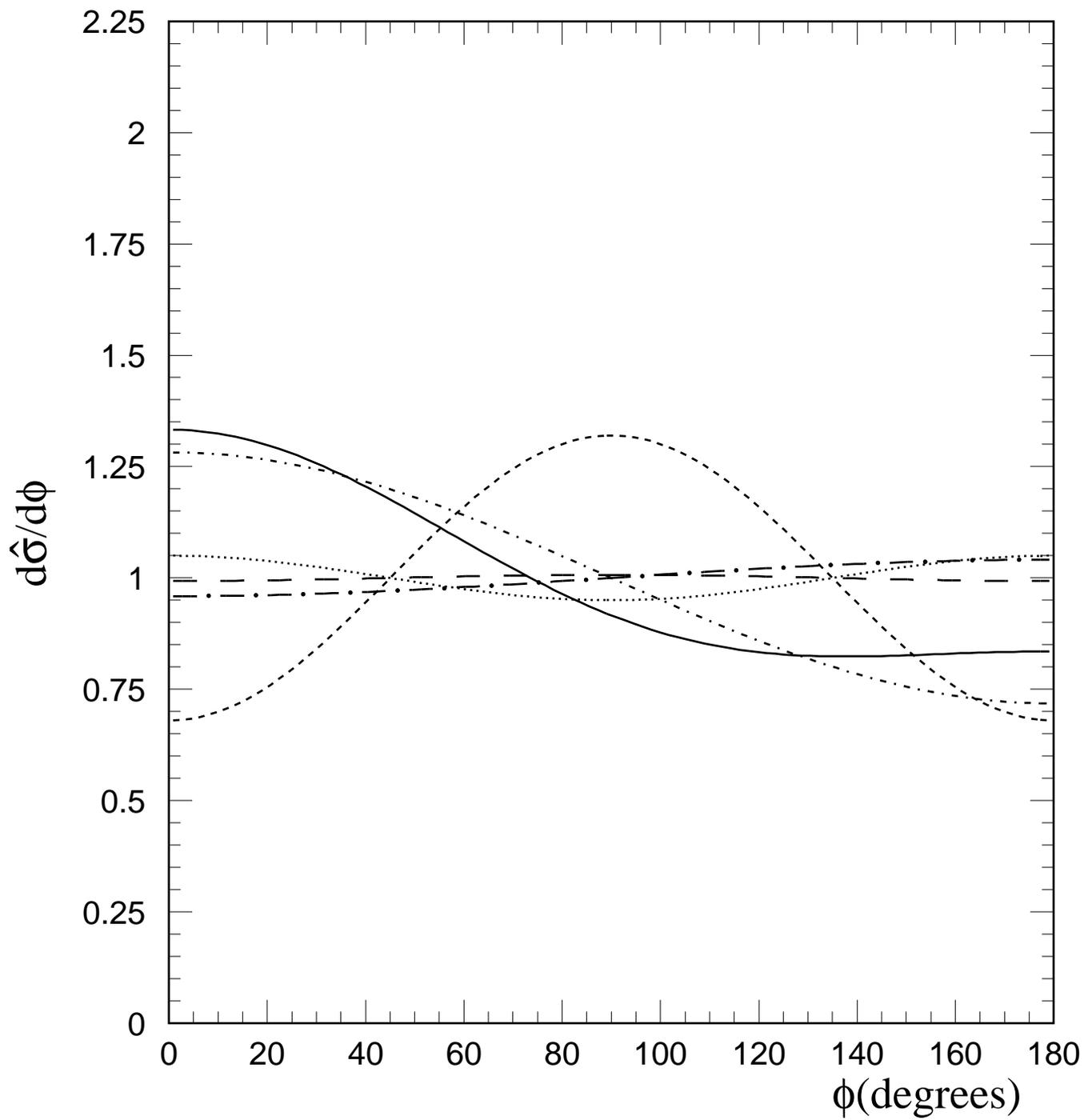
\begin{figure}
\caption{
Azimuthal correlations
($d\hat\sigma/d\Phi $) computed in configuration 4,
under the conditions defined in the text, for the reaction
$e \, e' \to e \, e' \, \mu^+ \mu^- $
at a beam energy of 5 GeV.
Solid line : $\Phi = \varphi_a $ ;
dotted line :  $\Phi = \varphi - \varphi_a $ ;
long-dashed line :  $\Phi = \varphi $ ;
short-dashed line : $\Phi  = 2 \varphi_a - \varphi $ ;
long-dashed/dotted line : $\Phi  = 2 \varphi -\varphi_a $ ;
short-dashed/dotted line :  $\Phi = 2 \varphi - 3 \varphi_a $.
}
\end{figure}
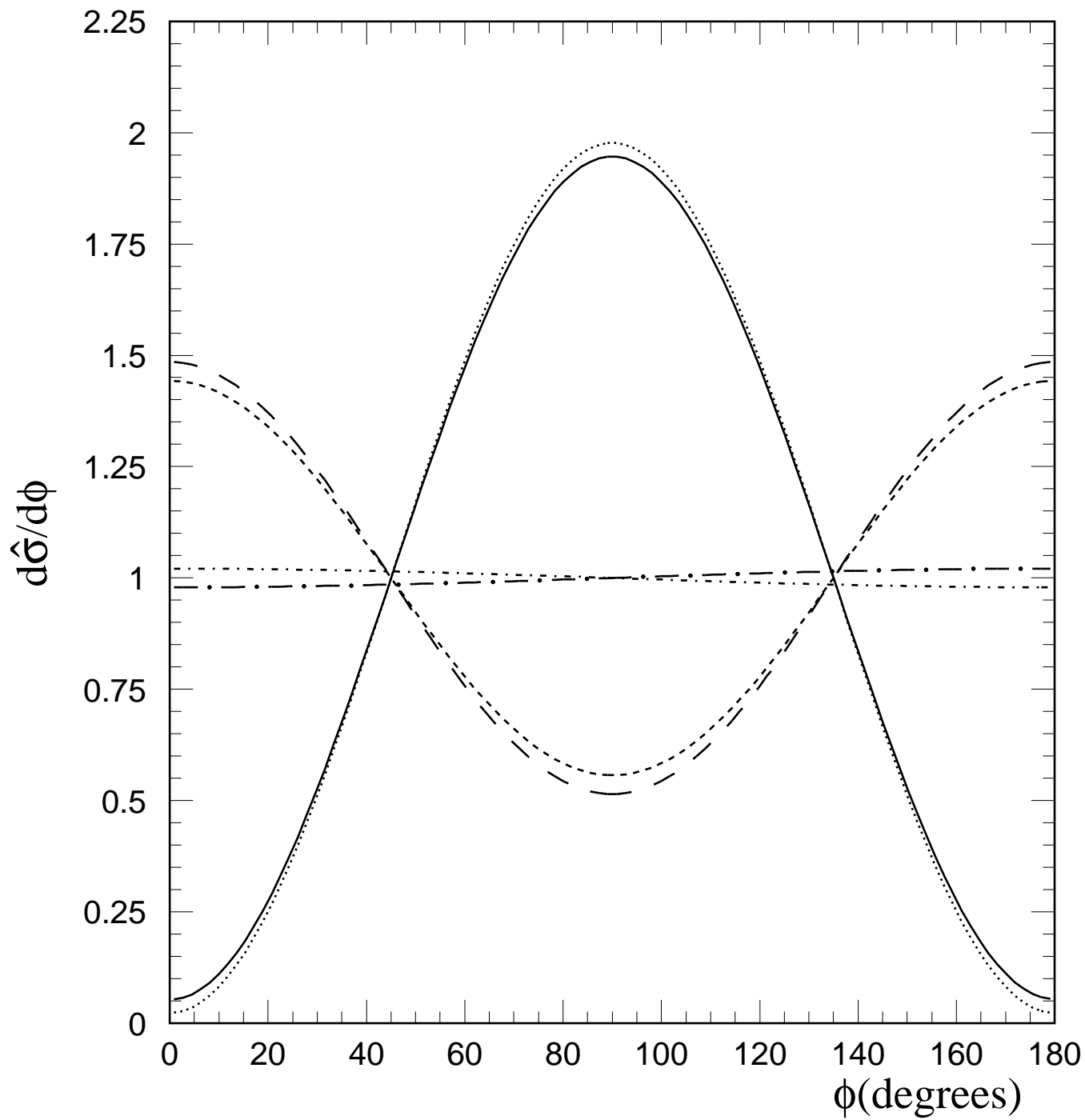
\begin{figure}
\caption{
Same as Fig. 2, for the reaction
$ e \, e' \to e \, e' \, \pi^+ \, \pi^- $
computed in the Brodsky-Lepage model \protect\cite{brodsky},
as extended by Gunion et al. \protect\cite{gunion},
using the Chernyak-Zhitnitsky wave function \protect\cite{chernyak}.
}
\end{figure}
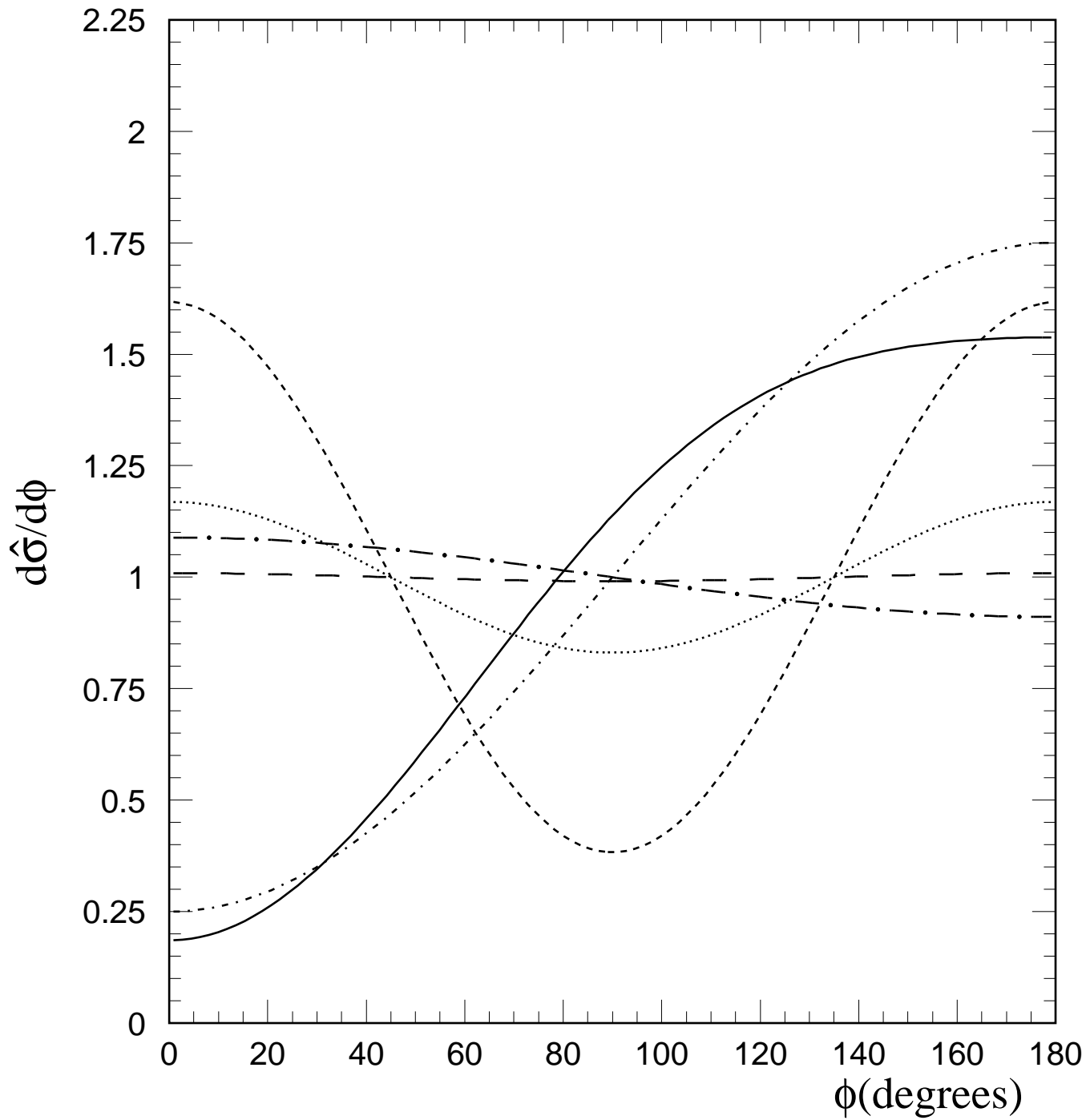
\begin{figure}
\caption{\protect
Same as Fig. 2, for the reaction
$e \, e' \to e \, e' \, \pi^+ \pi^- $
computed in the finite-size model of Poppe \protect\cite{poppe}.
}
\end{figure}
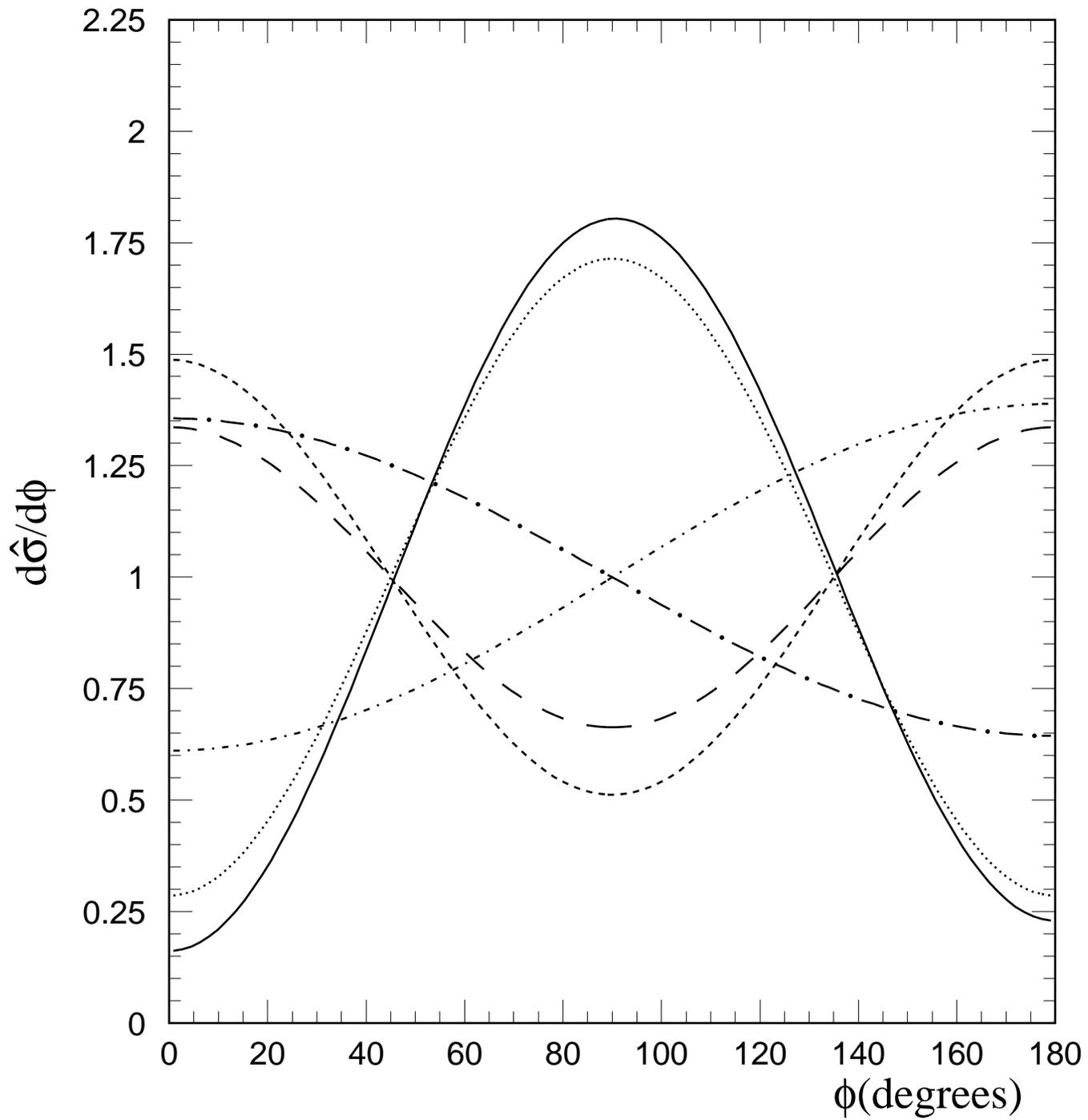
\begin{figure}
\caption{\protect
Same as Fig. 2, for the reaction
$e \, e' \to e \, e' \, \pi^0 \pi^0 $
computed in the Brodsky-Lepage model \protect\cite{brodsky},
as extended by Gunion et al. \protect\cite{gunion},
using the Chernyak Zhitnitsky wave function \protect\cite{chernyak}.
}
\end{figure}
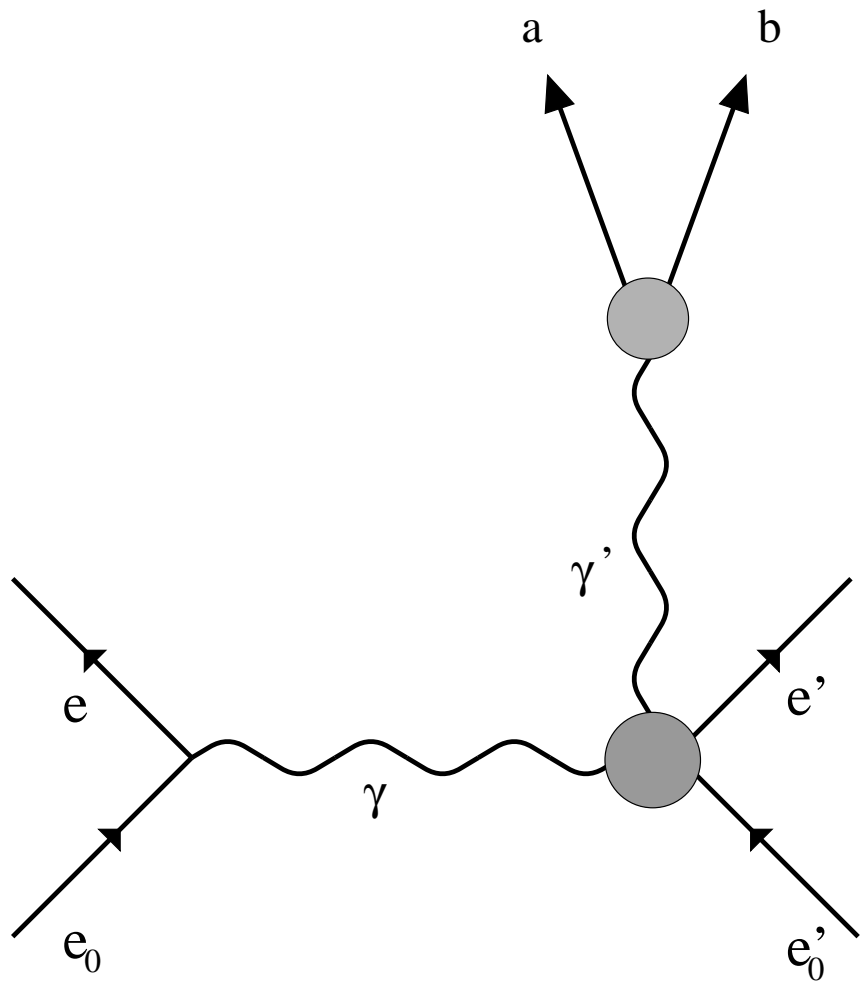
\begin{figure}
\caption{
Feynman diagram for the reaction
$ e \, e' \to e \, e' \, a \, b $
involving the exchange of one spacelike and one timelike photon.
}
\end{figure}
\begin{table}
\caption{
Comparison between the integrated cross sections, computed in
configuration 4 under the conditions defined in the text,
of the reactions
$ e \, e' \to e \, e' \, a \, b $
as described by the Feynman diagrams of Fig. 1 (signal)
and of Fig. 6 (background) respectively.
}
\vspace{0.5cm}
\begin{tabular}{cddd}
$ a\, b $ & $E_0$ & $\sigma_{\text{signal}}$
& $\sigma_{\text{background}}$\cr
\ & (GeV) & $(10^{-40} \text{cm}^2$) & $(10^{-40}\text{cm}^2$)\cr
\tableline
$\mu^+ \,\mu^- $ & 5. & 24 020. & 88.6    \cr
$\pi^+ \,\pi^- $ & 5. & \tablenotemark[1]142. & 0.215\cr
$\pi^+ \,\pi^- $ & 5. & \tablenotemark[2]249. & 0.215 \cr
$\pi^0 \,\pi^0 $ & 5. &    20. & --- \cr
$\mu^+ \,\mu^- $ & 100. & 222. & 0.283\cr
\tablenotetext[1]{Brodsky-Lepage\protect\cite{brodsky}}
\tablenotetext[2]{Poppe\protect\cite{poppe}}
\end{tabular}
\end{table}

\begin{thebibliography}{99}
\bibitem{note}
              This may not be verified in some particular cases, e.g.
              when transverse-photon interactions are inhibited by
              a selection rule (as the Landau-Yang rule for
              spin 1 resonance production) or possibly if a certain inner
              scale comes into the picture (see. Ref.~\cite{budnev2}).
\end{thebibliography}
\end{document}